# Lithography-free control of the position of single walled carbon nanotubes on a substrate by focused ion beam induced deposition of catalyst and chemical vapor deposition

El-Hadi S. Sadki[1]*, Ryo Matsumoto[2], Hiroyuki Takeya[2], and Yoshihiko Takano[2]*

[1]*Physics Department, College of Science, UAE University, Al Ain, UAE*

[2]*International Center for Materials Nanoarchitectonics (MANA), National Institute for Materials Science (NIMS), Tsukuba 305-0047, Japan*

E-mail: e_sadki@uaeu.ac.ae, takano.yoshihiko@nims.go.jp

We introduce a novel nanofabrication technique to directly deposit catalyst pads for the chemical vapor deposition synthesis of single-walled carbon nanotubes (SWCNTs) at any desired position on a substrate by Gallium focused ion beam (FIB) induced deposition of silicon oxide thin films from the metalorganic Tetraethyl orthosilicate (TEOS) precursor. A high resolution in the positioning of the SWCNTs is naturally achieved as the imaging and deposition by FIB are conducted concurrently in situ at the same selected point on the substrate. This technique has substantial advantages over the current state-of-the-art methods that are based on complex and multistep lithography processes.

Single-walled carbon nanotubes (SWCNTs), with their superior electronic and mechanical properties,[1-3] are promising nanoscale materials for prospective nanodevice applications in electronics[4-7] and molecular nano-sensing.[8,9] However, there are still many technical nanofabrication challenges hindering their implementation in commercial electronics products.[5] One of main experimental challenges is the control of their position on a substrate.[4] The standard nanofabrication method used to achieve this task requires several steps. First, alignment marks are made on the substrate by electron beam lithography, metal evaporation or sputtering, and lift-off. Then the nanotubes are either dispersed from a solution on the substrate, or are synthesized by chemical vapor deposition (CVD), after another electron beam lithography, evaporation or sputtering of catalyst films for the CVD process, and lift-off are conducted. The alignment marks are used as reference points to locate the nanotubes on the substrate for further electron beam lithography, evaporation or sputtering, and lift-off processes to make electrical contacts onto the nanotubes for either electrical transport properties characterization or device fabrication. This is clearly a lengthy and complex multistep nanofabrication procedure. Furthermore, the resolution in the positioning of the nanotubes on the substrate is limited by the hardware and software performance of the used electron beam lithography system in resolving the locations of the alignment marks.

In this paper, we propose a novel nanofabrication method for the positioning of SWCNTs on a substrate without the use of lithography and alignment procedures. It consists of using a gallium focused ion beam (FIB) to induce the deposition of silicon oxide thin films pads from the precursor Tetraethyl orthosilicate (TEOS) on specific locations on the substrate,[10] and the deposited films act as catalysts for the CVD synthesis of the SWNCTs at exactly the deposition locations. It is analogous to directly "write" or "draw" the catalyst films pads on the substrate, with the TEOS molecules acting as the "ink". In this method, there is no need for alignment procedures as the imaging and film catalyst deposition are conducted concurrently in situ at the same selected point on the substrate by the FIB. Furthermore, a high resolution in the positioning of the films is achieved, as it is limited only by the diameter of the ion beam used, which is in the order of ten nanometers only.[11] It is noted in a paper by Pander et al.[12] that FIB was used to pattern catalyst films (pre-deposited by sputtering) on a substrate leading to the CVD synthesis of vertically-aligned carbon nanotubes bundles or "forests". Although, their patterning method requires extra steps than ours (e.g. sputtering of the films in a different facility), it is more suited for applications that require vertically-aligned and high-density carbon nanotubes, such as superlenses, antennas, and thermal metamaterials.[12]

In this work, the synthesized SWCNTs are characterized by scanning electron microscopy (SEM), Atomic force microscopy (AFM), and Raman spectroscopy. The composition and morphology of the catalyst thin films are investigated by Energy-dispersive X-ray spectroscopy (EDS) and AFM, respectively. The yield of the synthesized SWCNTs is studied as function of the dimensions of the deposited thin films pads, and the optimum values to obtain on average only one SWCNT grown from a thin film pad are found.

Figure 1 shows a schematic diagram of the nanofabrication process. First, in a commercial FIB system (SMI9800SE, SII NanoTechnology Inc), a gallium focused ion beam is used to decompose TEOS molecules at specific locations on a silicon oxide substrate, and induce the deposition of silicon oxide thin films pads on these locations. The FIB induced deposition is conducted with a gallium ion beam energy of 30 KeV, and under a pressure of $3 \times 10^{-3}$ Pa. The areas of the deposited thin film pads used in this study are 2 $\mu$m × 2 $\mu$m, 5 $\mu$m × 5 $\mu$m, and 5 $\mu$m × 10 $\mu$m, with thickness ranging from 3 to 700 nm. After the films' deposition, carbon nanotubes are synthesized by thermal CVD in a quartz tube furnace (Asahi Rika Corp.). The CVD recipe is as follows: The furnace is pre-heated to 900 °C in air. The substrate is inserted directly from room temperature inside the furnace, and annealed in a flow of $O_2$ (100 sccm) and Ar (400 sccm) mixture for 30 min at 900 °C. Next, the furnace is evacuated down to a pressure of 1 Pa, then Ar is injected inside the furnace until the pressure reaches 1 atm, and followed by a flow of $H_2$ (200 sccm) for 1 min. Finally, the synthesis step of the carbon nanotubes from the thin films pads is conducted under the co-flow of $CH_4$ (300 sccm) and $H_2$ (200 sccm) for 20 mins. After the synthesis, the furnace is cooled down to room temperature in $H_2$ flow.

Figure 2 presents SEM images of the synthesized carbon nanotubes (SU-70, Hitachi). From a film pad deposited over an area of 5 $\mu$m × 10 $\mu$m, and a thickness of 10 nm, tens of carbon nanotubes are grown, with lengths up to several microns (Fig. 2(a)). On the other hand, only one and long (several tens of microns) carbon nanotube is obtained from a 2 $\mu$m × 2 $\mu$m pad, with the same thickness of 10 nm (Fig. 2(b)).

To elucidate the nature of the synthesized carbon nanotubes, AFM (Nanocute, Hitachi), and Raman spectroscopy (InVia, Renishaw) are used. An AFM topography image, conducted in the cyclic contact AC mode, of a typical synthesized carbon nanotube is shown in Fig. 3(a). From its height profile (Fig. 3(b)), the estimated diameter of the nanotube is about 1.7 nm. The Raman spectroscopy measurement of a synthesized SWCNT is shown in Figs. 3(c) and 3(d). The G-band peak (detected with a 532 nm wavelength laser) appears as expected at around 1580 cm$^{-1}$, and there is no apparent D-band peak, which indicates that the synthesized SWCNT is nearly defect-free.[13] The RBM peak is detected (with a 633 nm wavelength laser) at the frequency $\omega_{RBM}$ = 194 cm$^{-1}$ that corresponds to a SWCNT's diameter of $d$ = 1.3 nm according to the formula $d = 248/\omega_{RBM}$ (nm).[13] Furthermore, to define the chirality of this SWCNT, a Kataura plot[13,14] analysis is used as shown in Fig. 3(e). The laser's energy of 1.96 eV (633 nm wavelength) should be in resonance (with a typical resonance window of 50 meV) with one of the optical transitions of the SWCNT, which, in this case, corresponds to the metallic SWCNT, with the chirality (12,6) and diameter of 1.26 nm. This is indeed in excellent agreement with the calculated value of 1.3 nm from the RBM frequency. All the synthesized SWCNTs in this work have typical diameters between 1 to 3 nm.

It is widely accepted, and confirmed both experimentally and theoretically, that the main mechanism for the synthesis of carbon nanotubes by CVD is the so-called vapor-liquid-solid (VLS) mechanism, in which catalyst nanoparticles acts as "seeds" for the growth of carbon nanotubes.[15,16] Figure 4(a) shows an AFM topography image of a SWCNT emerging from a cluster of nanoparticles originating from a deposited film of 10 nm in thickness. The diameter of the nanoparticles is comparable to that of the synthesized SWCNTs (Fig. 4(b)), which proves that they can act as catalysts for the synthesis of the SWNCTs via the VLS mechanism. However, the most common catalyst nanoparticles used for the synthesis of carbon nanotubes by CVD are made of the transition metals Fe, Co, or Ni,[15-17] which are absent in this current work. To explore the materials composition of the catalyst nanoparticles obtained in our experiments, EDS analysis (Genesis APEX4, EDAX/AMETEK Inc) in the SEM facility is used. Figure 4(c) shows typical EDS spectra, conducted with electron beam energy of 5 KeV, of a 10 nm thickness film after FIB deposition, and on the resulting nanoparticles from the same film after CVD. After the FIB deposition, in addition to Si and O, the film contains significant amounts of gallium and carbon elements. Clearly Ga originates from the FIB, and carbon from the TEOS precursor or/and carbon contamination in the FIB system chamber. However, after CVD, the Ga and C peaks are absent, and only that of Si and O remain. This is explained by the fact that during the used CVD process carbon is burned away by the oxygen annealing, and gallium is evaporated at the high operating temperature (900 °C). Furthermore, it is found that if the CVD recipe does not include an annealing step in O$_2$, carbon nanotubes are not synthesized (See supplementary data). Hence, the removal of carbon by annealing before the synthesis step is crucial.

From the above analysis, it is concluded that the catalyst nanoparticles in our work are made of silicon oxide or of the general non-stoichiometric form $SiO_x$. Indeed, several previous reports have shown that $SiO_x$ nanoparticles can be used as catalyst for the synthesis of SWCNTs by CVD.[18-22] Moreover, it has been recently reported by Zhang et al.[22] that SWCNTs could be selectively grown with metallic or semiconducting properties by tuning the size of the catalyst $SiO_x$ nanoparticles and the CVD recipe. It is noted that our CVD recipe is similar to one of the reported recipes in Ref. 22. However, in contrast to their results, we could not obtain SWCNTs without the $O_2$ annealing step. This could be explained by the expected low content of carbon in their $SiO_x$ films, which were made by sputtering, in contrast with our FIB deposited films.

Different FIB deposited thin films dimensions and CVD recipes were tried in order to optimize the control of the yield of the synthesized SWCNTs (See supplementary data). Specifically, to obtain just one SWCNT on average the optimum initial film dimensions are 2 $\mu$m × 2 $\mu$m, and 10 nm, for the area and thickness, respectively. If either the area or thickness is increased, it is found that the yield (i.e. the number) of the synthesized SWCNTs increases. Any area or thickness smaller than the optimized values yields to no carbon nanotubes synthesis in general (See supplementary data).

For applications, our suggested nanofabrication technique will be very useful when the precise positioning of a SWCNT at or near a specific point on a substrate is required. Examples are the synthesis of a SWCNT directly on metal electrodes for electrical transport characterization,[6,23-25] and the positioning of a SWCNT over a nanopore for biological molecules sensing.[26] On the other hand, if a high resolution in the positioning of individual SWCNTs is not required and/or a large area on a substrate needs to be patterned by SWCNTs, then conventional lithography techniques would be more effective in terms of processing time and cost. Furthermore, for nanodevice applications, our method can be dramatically improved by simultaneously controlling the alignment of the synthesized SWCNTs by either substrate-oriented growth[25,27,28] or gas flow-rate CVD.[29,30] Another interesting idea is to explore different precursors for the FIB induced deposition that might lead to more efficient catalyst nanoparticles for the CVD synthesis of carbon nanotubes.

In summary, we introduce a novel nanofabrication technique for controlling the position of single walled carbon nanotubes on a substrate, without the use of lithography methods. It consists of the deposition of silicon oxide nanoparticles thin films by gallium focused ion beam at desired points on a substrate, with the nanoparticles acting as catalysts for the CVD synthesis of single walled carbon nanotubes at these specific points. The optimum area and thickness of the thin film catalyst to yield to the synthesis of only one nanotube on average is determined. The proposed technique is an important contribution to the field of carbon nanotubes synthesis and nanofabrication.

**Supplementary Data**

Chemical vapor deposition (CVD) without the oxygen annealing step, and effect of the films initial thickness on the yield of the synthesized carbon nanotubes (PDF).

**Acknowledgments**

ESS acknowledges the financial support from NIMS. ESS is grateful to H. Alawadhi, K. Daoudi, and M. Shameer, from the Center of Advanced Materials Research at the University of Sharjah, for their support with Raman spectroscopy measurements

**Figure Captions**

**Fig. 1.** Schematic diagram of the nanofabrication process: (a) Gallium focused ion beam (FIB) decomposes TEOS precursor molecules over specific positions on a substrate, where carbon nanotubes are intended to be located. (b) This leads to the deposition of silicon oxide thin films pads at these locations. (c) Using chemical vapor deposition (CVD), single walled carbon nanotubes are synthesized from silicon oxide nanoparticles that originate from the deposited thin films.

**Fig. 2.** Scanning electron microscopy (SEM) images of single walled carbon nanotubes synthesized from a 5 $\mu$m × 10 $\mu$m (a) and 2 $\mu$m × 2 $\mu$m (b) silicon oxide thin films of thickness 10 nm. From the 2 $\mu$m × 2 $\mu$m film (b), only one carbon nanotube is synthesized.

**Fig. 3.** (a) Atomic force microscopy (AFM) topography image of a synthesized single walled carbon nanotube. (b) The height profile along the blue line in (a) shows that the diameter of the nanotube is about 1.7 nm. (c) Raman spectrum of a single walled carbon nanotube that shows the G-band peak, and the expected position for the D-band peak (dotted vertical line). (d) Raman spectrum of a single walled carbon nanotube that shows the RBM peak at 194 cm-1 (The peak at 303 cm$^{-1}$ is due to the silicon substrate). (e) A Kataura plot of SWCNTs optical energy transitions versus nanotube's diameter showing the resonance region of the incident photons (from the laser), with a resonance window of 50 meV. The red circles and green squares represent metallic and semiconducting SWCNTs, respectively. The metallic SWCNT, with chirality (12,6), and diameter 1.26 nm that matches the RBM peak's position, is the only SWCNT that satisfies these conditions.

**Fig. 4.** (a) Atomic force microscopy (AFM) topography image of silicon oxide nanoparticles and a carbon nanotube after chemical vapor deposition (CVD) with a 10 nm thickness silicon oxide thin film. (b) The height profile along the blue line in (a) shows the presence of nanoparticles with diameters in the same range as the diameter of the synthesized carbon nanotubes (1 to 3 nm). (c) Energy-dispersive X-ray spectroscopy (EDS) spectrum of a silicon oxide thin film of thickness 10 nm before (Left) and after (Right) chemical vapor deposition (CVD). The data before CVD shows peaks corresponding to Gallium and Carbon elements. However, after CVD these peaks are absent.

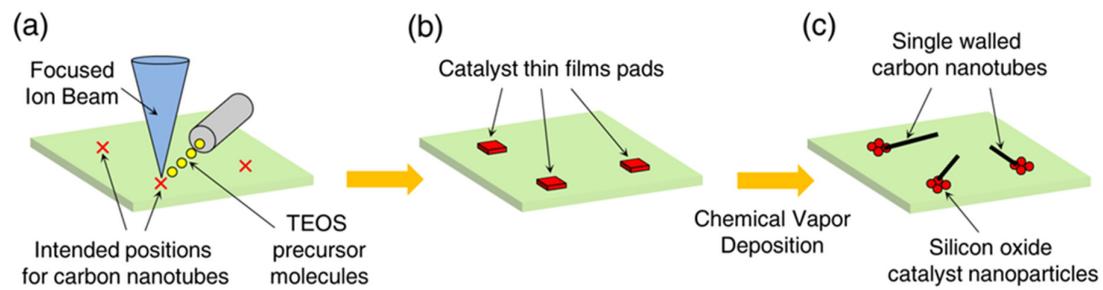

Fig. 1.

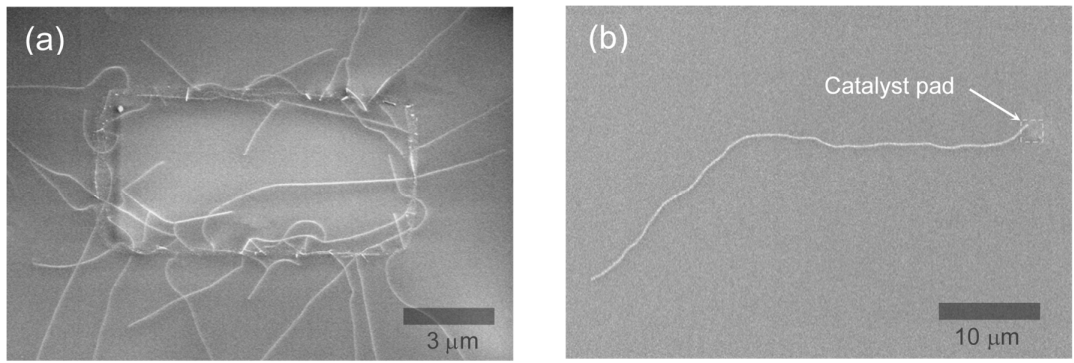

Fig. 2.

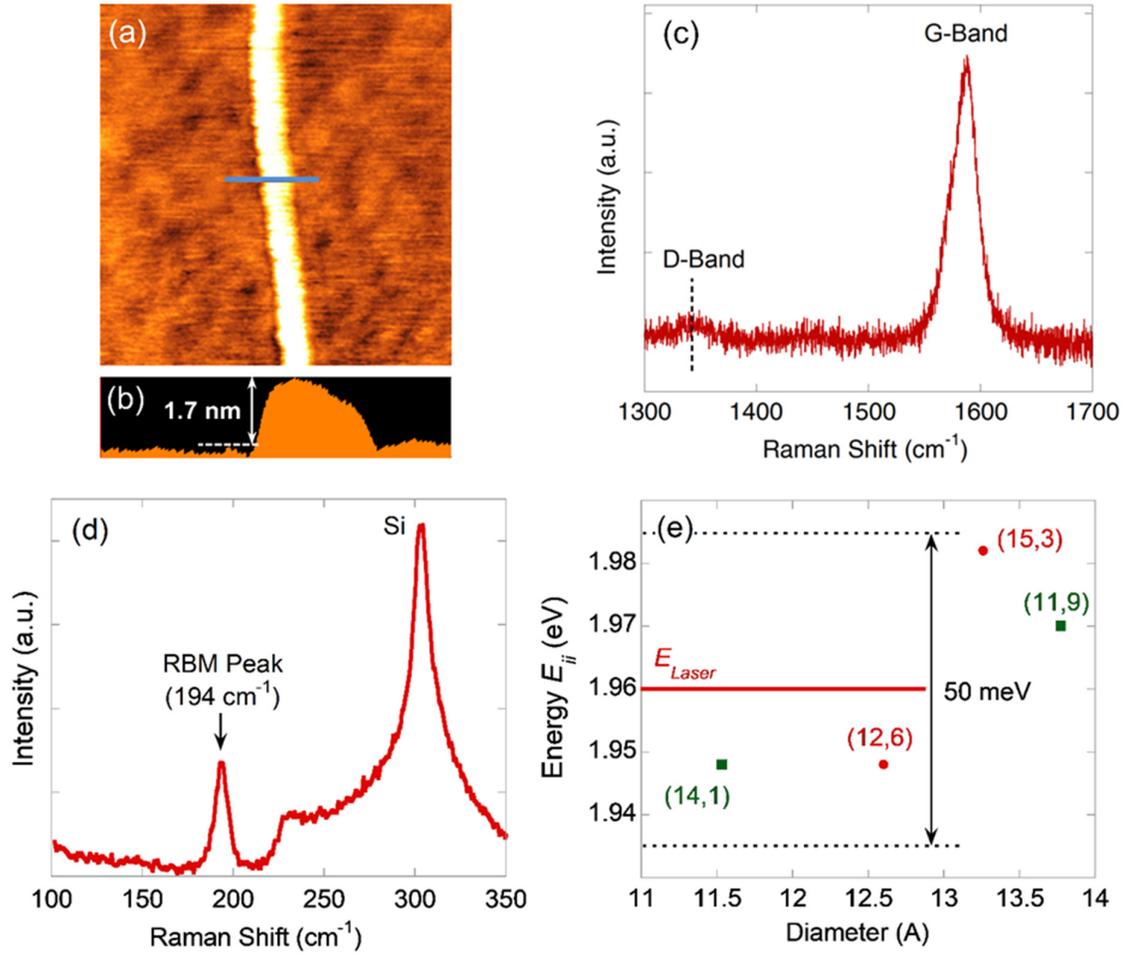

Fig. 3.

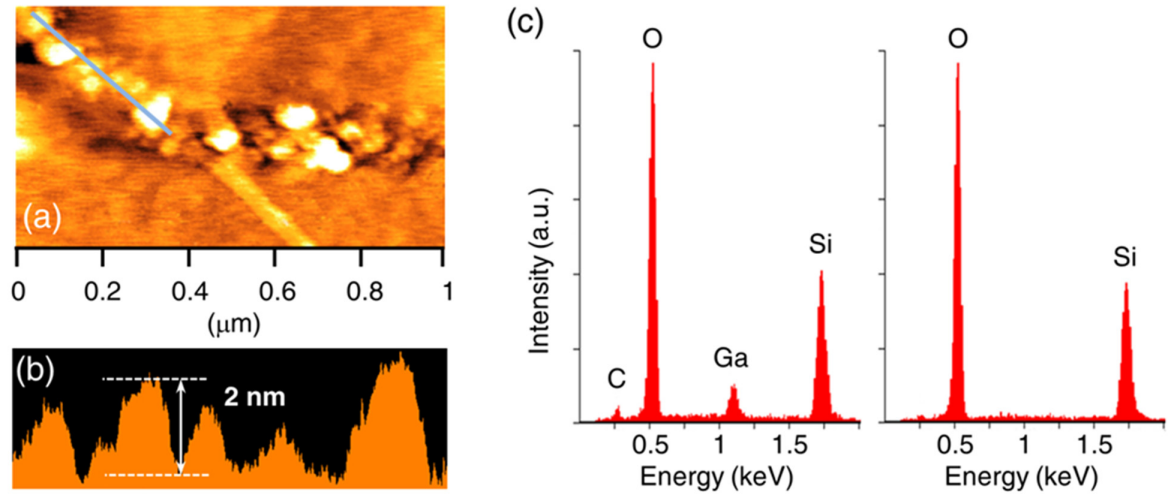

Fig. 4.

Supplementary Data for

# Lithography-free control of the position of single walled carbon nanotubes on a substrate by focused ion beam induced deposition of catalyst and chemical vapor deposition

*El Hadi S. Sadki, Ryo Matsumoto, Hiroyuki Takeya, and Yoshihiko Takano*

International Center for Materials Nanoarchitectonics (MANA), National Institute for Materials

Science (NIMS), Tsukuba 305-0047, Japan

**1- Chemical Vapor Deposition (CVD) without the Oxygen annealing step**

A CVD recipe without the oxygen annealing step was also tested on the FIB deposited films. Its details are shown in Figure S1. The samples were annealed in Argon flow from room temperature to 900 ºC, and then the standard co-flow of $CH_4$ and $H_2$ was introduced for 20 mins. As mentioned in the main text, no carbon nanotubes were obtained from using this recipe.

Since the only main difference between the recipe that successfully led to SWCNTs growth (as described in the main text) and the latter recipe is the oxygen annealing step, it is concluded that this step is crucial for SWCNTs synthesis.

To elucidate the exact role of oxygen step in the CVD recipe, EDS analysis was conducted before and after CVD without oxygen annealing. The results are shown in Figure S2. It is clear from the results that carbon remains in the films after the CVD process. As we have concluded in the main text, the presence of carbon in the films before the methane/hydrogen step prevents the synthesis of the SWCNTs as carbon "poison" the $SiO_x$ nanoparticles, and prevent them from absorbing carbon from methane according to the VLS mechanism.

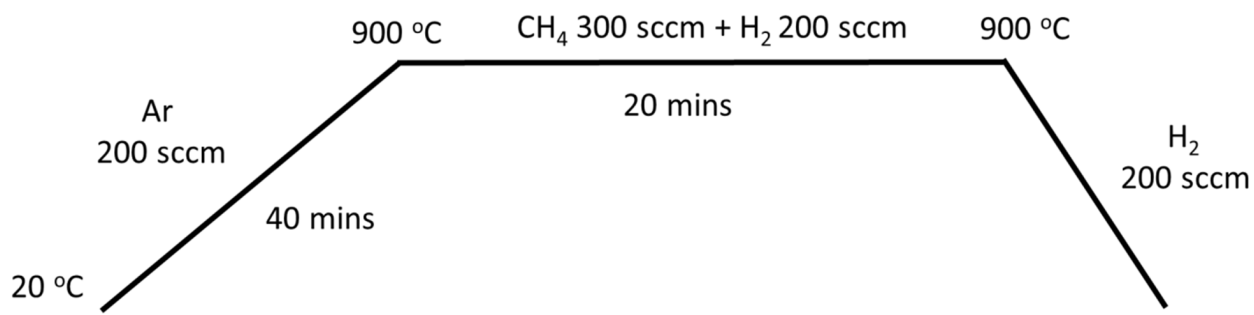

**Figure S1.** Chemical vapor deposition (CVD) recipe without oxygen annealing.

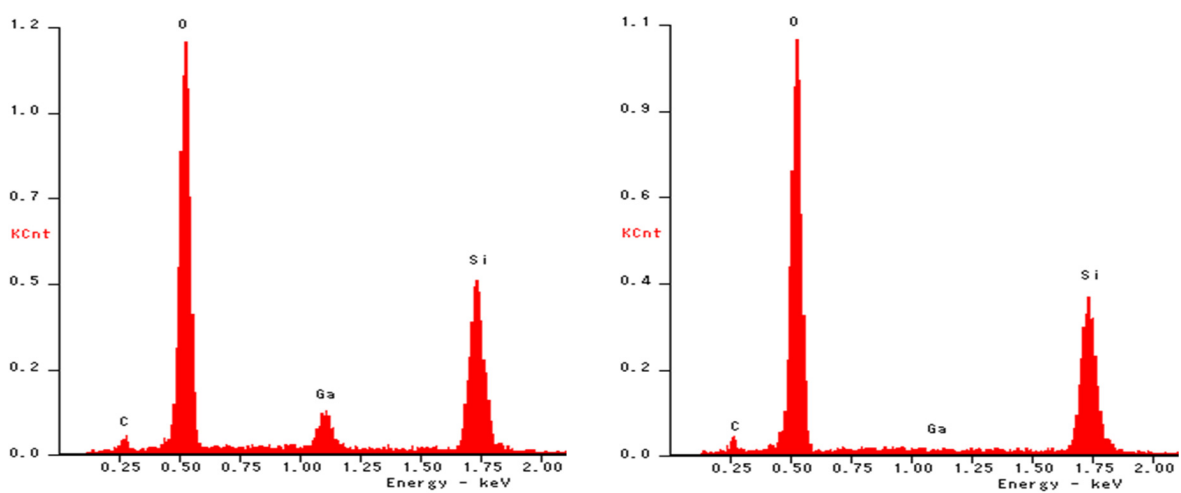

**Figure S2.** EDS spectra of a 10 nm thickness of a catalyst film before CVD (Left) and after CVD without oxygen annealing (Right). The carbon peak is present in both spectra, and Ga peak is absent after CVD.

**2- Effect of the Films initial thickness on the yield of the synthesized carbon nanotubes**

Figure S3 shows SEM images of FIB deposited films after the CVD process. With the same CVD recipe as in the main text of the manuscript, the yield of the nanotubes depends on the film's thickness, and increases with the film's thickness (Top image). An example is shown at the bottom left image of a film of thickness 10 nm and area of 5 $\mu$m x 5 $\mu$m: Two nanotubes were synthesized (One long and one short). However, the yield is also affected by the CVD's recipe, as shown at bottom right image, where a film of thickness 50 nm and area of 5 $\mu$m x 5 $\mu$m produced only two nanotubes (One long and one short), which were synthesized with a shortened oxygen annealing time during the CVD process.

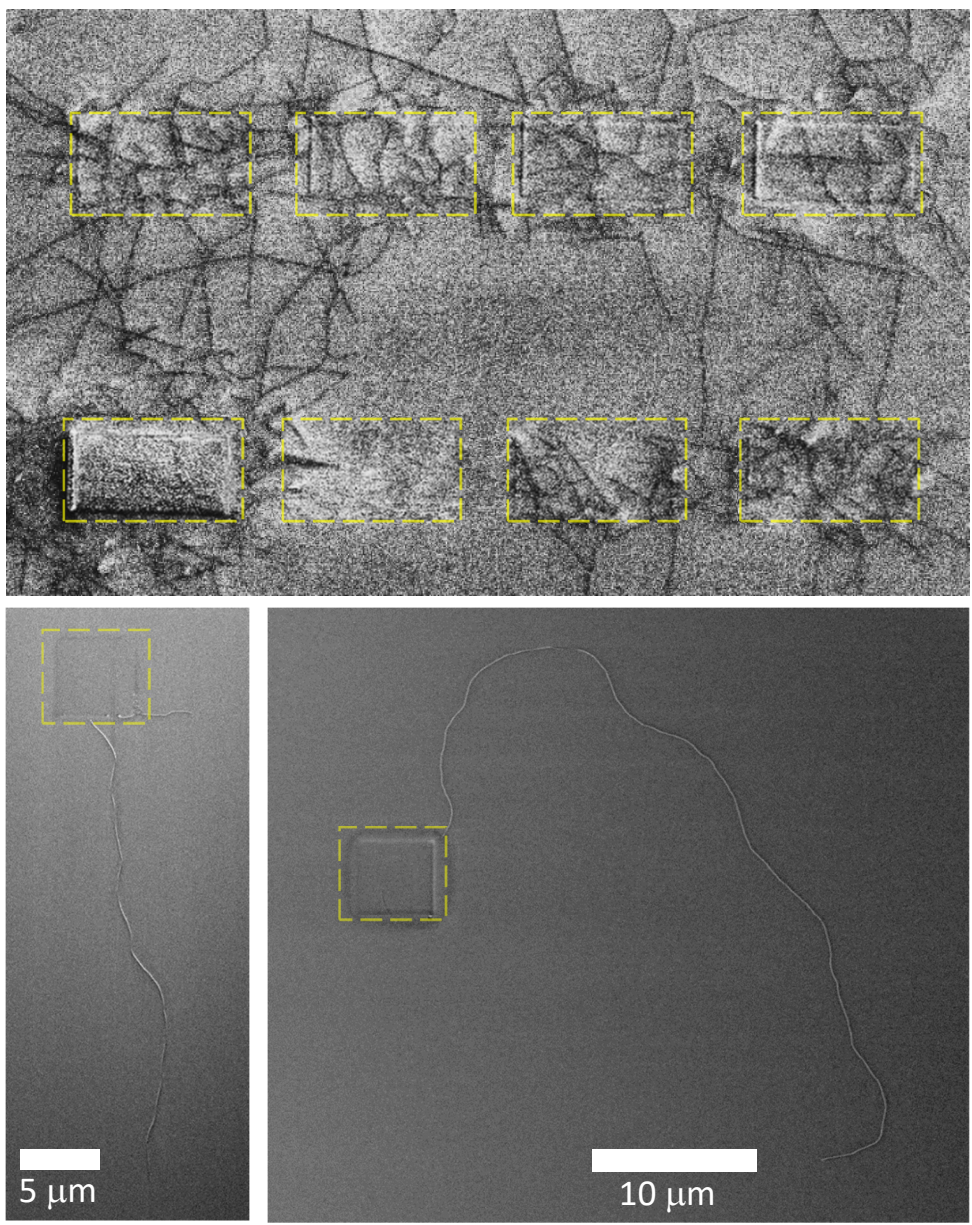

**Figure S3.** SEM images of FIB deposited films after CVD process. (Top) Array of eight films of area of 10 μm x 5 μm and thicknesses ranging from 3 nm to 500 nm. The dotted lines show the location of the films. The nanotubes appear as either dark or white lines due to strong charging effects from the silicon oxide substrate. (Bottom left) A film of thickness of 10 nm and area of 5 μm x 5 μm: Two nanotubes were synthesized (One long and one short). (Bottom right) A film of thickness of 50 nm and area of 5 μm x 5 μm: Two nanotubes (One long and one short) were synthesized with a shortened oxygen annealing time in the CVD process.

For film thickness less than 10 nm, with the CVD recipe described in the text, no nanotubes are produced. In Figure S4, a film's thickness of 5 nm produced no carbon nanotubes (Left). For a thickness well above that, for example a thickness of 500 nm (Right), a very high yield of carbon nanotubes is obtained.

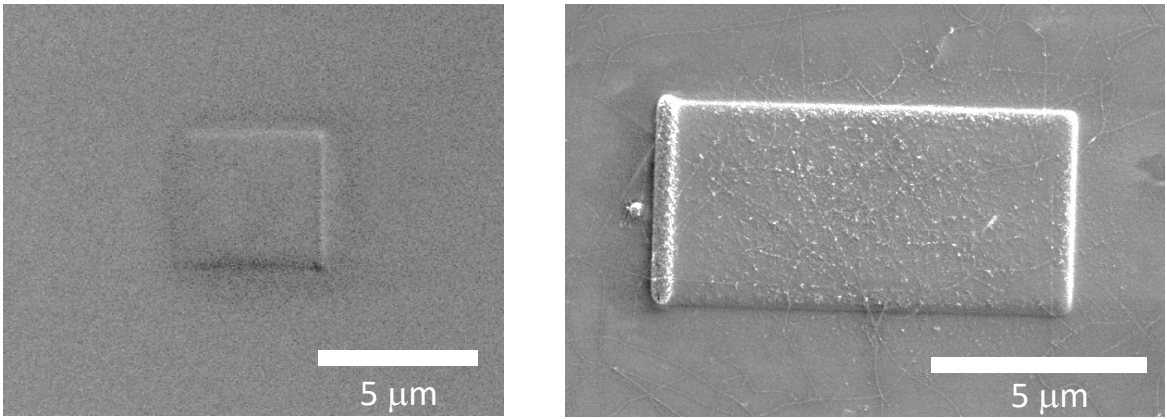

**Figure S4.** SEM images of FIB deposited films after CVD process. (Left) Initial film thickness of 5 nm and area of 5 $\mu$m x 5 $\mu$m: No carbon nanotubes were synthesized. (Right) A film thickness of 500 nm and area of 5 $\mu$m x 10 $\mu$m: A lot of nanotubes were synthesized.